\documentclass[aps,pre,twocolumn,superscriptaddress,citesort,longbibliography]{revtex4-2}
\usepackage{graphicx}
\usepackage{amsmath}

\usepackage{hyperref}
\usepackage{xcolor}
\definecolor{tab_blue}{HTML}{1F77B4}
\hypersetup{%
  breaklinks=true,
  colorlinks=true,
  linkcolor=tab_blue,
  filecolor=tab_blue,
  urlcolor=tab_blue,
  citecolor=tab_blue,
}

\begin{document}

\title{Extreme statistics and extreme events in dynamical models of turbulence}
\author{Xander M. de Wit}
\affiliation{Department of Applied Physics and Science Education, Eindhoven University of Technology, 5600 MB Eindhoven, Netherlands}
\author{Giulio Ortali}
\affiliation{Department of Applied Physics and Science Education, Eindhoven University of Technology, 5600 MB Eindhoven, Netherlands}
\affiliation{SISSA (International School for Advanced Studies), via Bonomea 265, I-34136 Trieste, Italy}
\author{Alessandro Corbetta}
\affiliation{Department of Applied Physics and Science Education, Eindhoven University of Technology, 5600 MB Eindhoven, Netherlands}
\author{Alexei A. Mailybaev}
\affiliation{Instituto de Matemática Pura e Aplicada-IMPA, Estrada Dona Castorina, 110 Jardim Botânico, Rio de Janeiro, RJ 22460-320, Brazil}
\author{Luca Biferale}
\affiliation{Department of Physics and INFN, University of Rome ``Tor Vergata'', Via della Ricerca Scientifica 1, 00133 Rome, Italy}
\author{Federico Toschi}
\affiliation{Department of Applied Physics and Science Education, Eindhoven University of Technology, 5600 MB Eindhoven, Netherlands}
\affiliation{CNR-IAC, I-00185 Rome, Italy}
\date{\today}

\begin{abstract}
We present a study of the intermittent properties of a shell model of turbulence with unprecedented statistics, about $\sim 10^7$ eddy turn over time, achieved thanks to an implementation on a large-scale parallel GPU factory. This allows us to quantify the inertial range anomalous scaling properties of the velocity fluctuations up to the $24$-th order moment. Through a careful assessment of the statistical and systematic uncertainties, we show that none of the phenomenological and theoretical models previously proposed in the literature to predict the anomalous power-law exponents in the inertial range is in agreement with our high-precision numerical measurements. We find that at asymptotically high order moments, the anomalous exponents tend towards a linear scaling, suggesting that extreme turbulent events are dominated by one leading singularity. We found that systematic corrections to scaling induced by the infrared and ultraviolet (viscous) cut-offs are the main limitations to precision for low-order moments, while high orders are mainly affected by the finite statistical samples. The unprecedentedly high fidelity numerical results reported in this work offer an ideal benchmark for the development of future theoretical models of intermittency in dynamical systems for either extreme events (high-order moments) or typical fluctuations (low-order moments). For the latter, we show that we achieve a precision in the determination of the inertial range scaling exponents of the order of one part over ten thousand (5th significant digit), which must be considered a record for out-of-equilibrium fluid-mechanics systems and models.
\end{abstract}
\maketitle	

\section{Introduction} 
Turbulence, manifested in the vast majority of all natural and industrial flows around us, has remained among the most elusive problems in modern day classical physics. In spite of centuries of research, we still lack a complete theoretical description of the dynamics that is encompassed in turbulent flows. One of the hallmarks of 3D turbulence is the existence of an inertial range of scales limited in the infrared region by the forcing mechanism and the ultraviolet by the viscous dissipation. This inertial range is found to be universal, in the sense that it is largely independent of the detailed nature of the forcing or the dissipation mechanisms \cite{Frisch1995,benzi2023lectures}. Contrary to the original self-similarity hypothesis by Kolmogorov \cite{Kolmogorov1941}, the distribution of velocity fluctuations in the inertial range is observed to be scale-dependent, developing fatter tails at decreasing length scales (increasing wavenumbers). This is a signature of intermittency, the emergence of rare but anomalously extreme events that develop in the inertial range \cite{Frisch1995,benzi2023lectures}.

The analytical calculation of the intermittent anomalous scaling exponents is the {\it Holy Grail} of turbulence and theoreticians have proposed many different approaches and solutions, with no success so far (see below), due to the exceptional theoretical difficulties in dealing with scaling properties in strongly out-of-equilibrium systems. Another factor hindering further developments is connected to the limited accuracy with which exponents can be measured from numerical simulations in Navier-Stokes equations, due to the exceptional computational needs to manage an extended inertial range with long temporal integration \cite{Boratav1997,Gotoh2002,Ishihara2009,Benzi2010,Iyer2020,Buaria2023}. Experimental data are also difficult to exploit, because of the unavoidable presence of systematic non-homogeneous and anisotropic effects which limits the achievable accuracy \cite{Anselmet1984,Water1999,Shen2002,Saw2018}.

An alternative route to gain insights into turbulent dynamics lies in using reduced order modeling of turbulence. Historically, a valuable approach has proven to be what is referred to as shell models. In these systems, instead of resolving the full 3D space, one only models the energy that is contained in shells of logarithmically spaced wavenumbers \cite{Biferale2003}. With this broad class of models, intermittency with different and, in some cases, even controllable scaling properties can be studied \cite{mailybaev2021solvable}, making it a perfect model for studying the concept of anomalous scaling in the most general terms. In this work, we focus on a particular instance of the shell model that most closely matches Navier-Stokes turbulence, conserving the equivalent energy and helicity \cite{Lvov1998}. In spite of their very strong reduction of degrees of freedom, shell models have been able to reproduce much of the phenomenology of full Navier-Stokes turbulence. This includes the presence of anomalous intermittent fluctuations, the existence of a dissipative anomaly and the development of extreme non-Gaussian events \cite{Biferale2003}. Although shell models cannot be expected to fully reproduce the statistics of Navier-Stokes turbulence at the quantitative level, shell models do offer an optimal test bed for the development of theories for anomalous scaling in out-of-equilibrium systems in general, and for turbulence, based on the Fourier-space approach, in particular.

\begin{figure*}
    \centering
    \includegraphics[width=\linewidth]{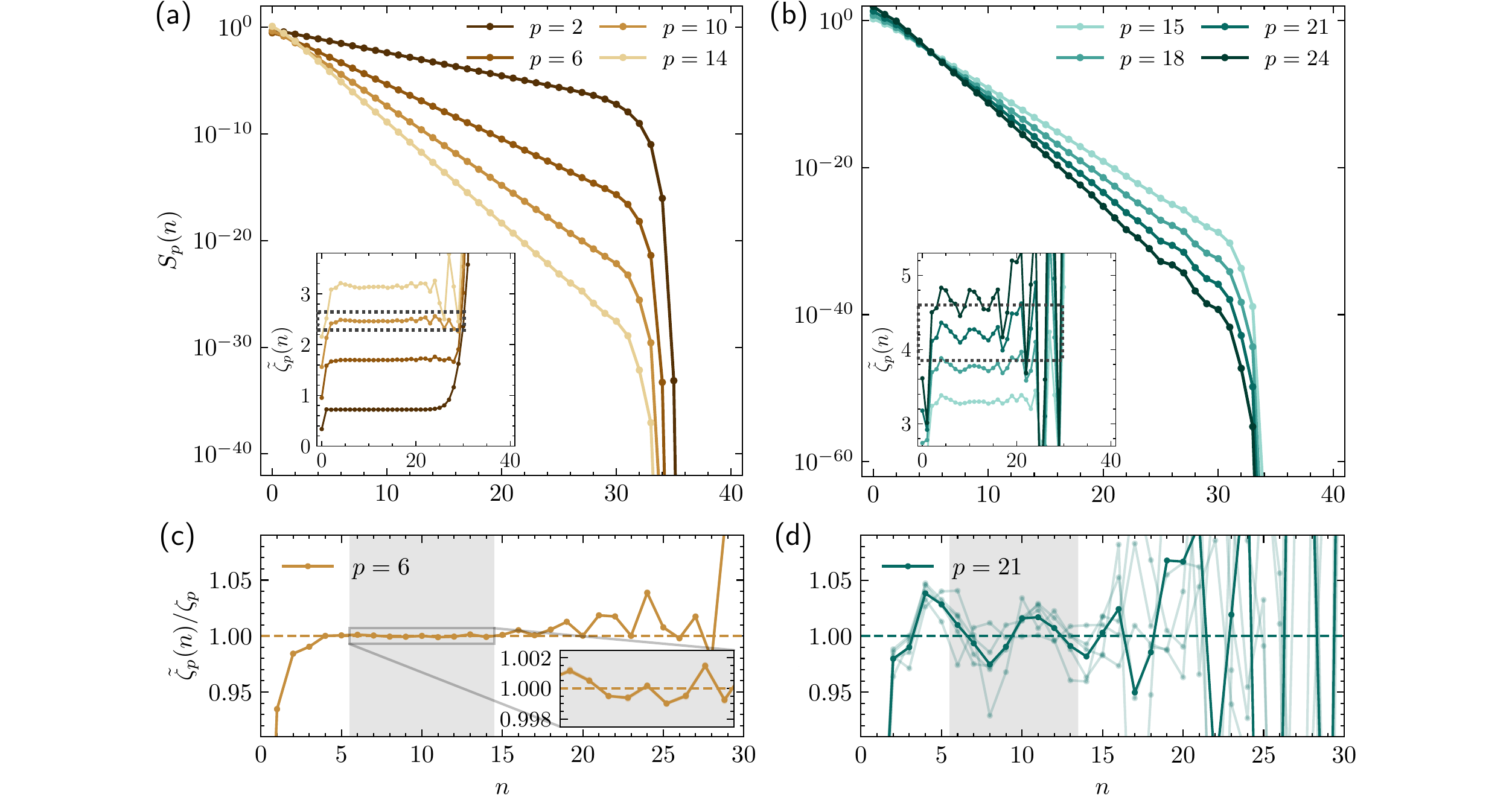}
    \caption{Panels a, b: Structure functions $S_p(n)$ and their local slopes $\tilde{\zeta}_p(n)$ (insets) for selected moments $p$. In the inertial range, the structure functions exhibit power-law scaling, similar to Navier-Stokes turbulence. Panels c, d: zoom on local slopes of the structure function $\tilde{\zeta}_p(n)$, normalized by their mean value $\zeta_p$ for $p=6$ (left) and $p=21$ (right). The vertical axis is strongly cropped to emphasize the fluctuations in the local slope. Solid lines represent the average over the full data, while the lighter shades represent averages over 1/5th subsamples of the data to reveal the statistical fluctuations. This shows that the lower moments (left) are dominated by fluctuations from the dissipative range that are systematic, being consistent across the subsamples, while the high moments (right) are instead dominated by statistical fluctuations. The selected inertial range is shaded in gray. Results are obtained over a total sample size of approximately $4\times10^7 \, T_L$.}
    \label{fig:fig_1}
\end{figure*}

A powerful interpretation of the intermittency in Navier-Stokes equations is provided by the Multifractal Model (MFM) as put forward by Parisi \& Frisch \cite{Parisi1985}, which can be justified from first-principles using the hidden scale invariance \cite{mailybaev2022hidden}. However, to fully prescribe the turbulent statistics, the MFM requires knowledge of the codimension, $C(h)=3-D(h)$, of the singularity spectrum $D(h)$. This is an unknown function that no one has been able to calculate from first-principles so far, highlighting once more the requirement for precise measurements of the turbulence statistics. 

In this work, we use GPU vectorization to capture very high order moment statistics with unprecedented precision for a popular shell model of turbulence. The resulting measurements of the anomalous inertial range scaling exponents $\zeta_p$ allow us to fully describe the intermittency statistics of the model, and compute the singularity spectrum $D(h)$ of the underlying multifractal. Thanks to the extreme precision achieved, we present reliable results up to moments of order $p=24$, and higher, never measured before either in Navier-Stokes turbulence or in other shell models studies. Our accuracy is high enough to exclude all theoretical proposals for the $\zeta_p$ curve published up to now and opens a clear new challenge to find a proper phenomenological or theoretical model for $\zeta_p$ in agreement with our data in the context of shell model turbulence.

\section{Numerical approach}
We employ the celebrated SABRA shell model \cite{Lvov1998}. It is governed by the dynamical equation for the shell velocity $u_n$ as function of time $t$ given by
\begin{equation}\label{eq:sabra}
\begin{aligned}
\frac{\mathrm{d} u_n}{\mathrm{d} t} = & i\left(k_{n+1} u_{n+2} u_{n+1}^* - \tfrac{1}{2} k_n u_{n+1} u_{n-1}^* \right. \\
& \left. +\tfrac{1}{2} k_{n-1} u_{n-1} u_{n-2} \right) -\nu k_n^2 u_n + f_n.
\end{aligned}
\end{equation}
This equation conserves energy $E=\sum_n |u_n|^2$ and helicity $H=\sum_n (-1)^n k_n |u_n|^2 $ in the inviscid unforced limit, in analogy to the full Navier-Stokes equation.
Here, log-spaced wavenumbers $k_n=k_0\lambda^n$ are indexed by shell number $n=0,1,...,N$, where we choose a spacing $\lambda=2$ and $k_0=1$. The asterisk $^*$ denotes complex conjugation. We choose a fixed large-scale forcing $f_n=(f_0,f_0/\sqrt{2},0,....,0)$. We pick a forcing amplitude $f_0=0.5$ and set the kinematic viscosity to $\nu=10^{-12}$, yielding a Reynolds number $\textrm{Re}\simeq 10^{12}$. This sets the dissipative Kolmogorov shell $N_\eta \simeq 30$ and the Kolmogorov time $\tau_\eta\simeq 1.8 \times 10^{-6}$. To accurately resolve the full dissipative range, we integrate up to shell $N=40$. The dynamical equation \eqref{eq:sabra} is integrated using a 4th order Runge-Kutta method with timestep $\Delta t = 10^{-8} \simeq 0.006 \tau_\eta$.

To be able to efficiently accumulate large statistics, we perform the shell model integration on GPU, integrating many ($M=4096$) independent replicas of the shell model in parallel on every employed GPU with statistically perturbed initial conditions. Implementation is done using the Google JAX package for efficient CPU and GPU handling \cite{JAX2018}. The shell model is continuously integrated on GPU, accumulating samples at a rate similar to the Kolmogorov time (sampling interval $t_{\textrm{samp}} = 10^{-5} \simeq 6\tau_\eta$). Accumulated samples then fill up a buffer that is passed in batches to CPU where the relevant summary statistics are computed in parallel asynchronously, clearing the GPU buffer.

To study the intermittency properties, we direct our attention to the structure function $S_p(n)$. In its most simple form, one can study the structure function of the velocity directly $\langle |u_n|^p \rangle$, but this particular structure function is known to be subject to significant structural period-3 oscillations, obfuscating its further analysis \cite{Lvov1998}. Therefore, we study the flux-based structure function instead, which combines consecutive shells as
\begin{equation}
    S_p(n) = \langle | F_n |^{p/3} \rangle,
\end{equation}
with \cite{Lvov1998}
\begin{equation}
F_n =\operatorname{Im}\left[\lambda u_n u_{n+1} u_{n+2}^*+\tfrac{1}{2} u_{n-1} u_n u_{n+1}^*\right].
\end{equation}

The angular brackets $\langle ... \rangle$ represent the ensemble average, which is taken by averaging over all time in the statistically steady state and over the different independent realizations.

In total, we collect around $3\times 10^{12}$ samples for every shell, which, in terms of the large Eddy turnover time $T_L=1/(k_0 \sqrt{\langle u_0^2\rangle})$, amounts to approximately $4\times 10^7 \, T_L$. To handle the massive stream of data produced by the raw sampling of the shell model ($\sim 2 \,\, \textrm{petabytes}$ in total), we retain only the running moments $S_p(n)$ as well as histograms of the flux $F_n$ itself. One needs to take caution when accumulating moments for very large data sets in order to avoid a loss of precision when the number of samples in the sum approaches the order of the machine precision. To avoid this problem, we recursively sum the moments in batches with a batch size $\mathcal{O}(10^3)$ to retain sufficient precision in the total sum.

\section{Inertial range scaling}
In the inertial range, the structure function, as in Navier-Stokes turbulence, follows a power law scaling with the wavenumber $k_n$ with a scaling exponent $\zeta_p$, i.e.
\begin{equation}
    S_p(n) \propto k_n^{-\zeta_p}.
\end{equation}
These scaling exponents $\zeta_p$ are universal and their dependence on the order of the moment $p$ fully describes the statistical intermittency properties of the inertial range.

To compute the scaling exponents $\zeta_p$ from the numerical data, we consider first the local slopes $\tilde{\zeta}_p(n)$, wavenumber by wavenumber, obtained as
\begin{equation}
    \tilde{\zeta}_p(n) = \frac{\log[S_p(n+1)]-\log[S_p(n)]}{\log[\lambda]}.
\end{equation}
Numerical results for the structure functions and their local slopes for some selected moments are provided in Fig.~\ref{fig:fig_1}.

The final estimate of the true scaling exponents $\zeta_p$ is then provided by the average of these local slopes over the inertial range $n_{\textrm{str}} \leq n < n_{\textrm{end}}$ as
\begin{equation}
    \zeta_p = \sum_{n_{\textrm{str}} \leq n < n_{\textrm{end}}} \frac{1}{n_{\textrm{end}} - n_{\textrm{str}}}  \tilde{\zeta}_p(n).
\end{equation}
We select the inertial range $6 \leq n < 15$ for moments \mbox{$p\leq 20$}. We gradually decrease the upper limit for moments $p>20$ down to $6\leq n < 11$ for the highest moment $p=24$ here considered, due to the finite sampling time as motivated in Appendix~\ref{app:hist}. The selected range of scales where the inertial range exponents have been evaluated is highlighted in gray in Fig.~\ref{fig:fig_1}.

However, crucially, the selection of the inertial range introduces a degree of arbitrariness and is therefore a source of systematic error. Due to the multifractal fluctuations there is not a unique dissipative wavenumber, and different moments will have different ultraviolet dissipative cut-offs \cite{Paladin1987,Frisch1991}. Furthermore, statistical fluctuations are also strongly dependent on the order of the moment and on the wavenumber, as well as the systematic fluctuations originating from the forcing and dissipative ranges. As a result, we have adopted the aforementioned empirical choice of the inertial range as a trade-off between on the one hand retaining a sufficient number of shells to make a reliable estimate of the overall scaling exponents, while on the other hand minimizing the influence of the different sources of error. As depicted in Fig.~\ref{fig:fig_1}(c,d), in particular the fluctuations coming from the ultraviolet travel far into the inertial range. The local slopes display a clear imprint of period-3-like oscillations coming from the dissipation range. These fluctuations are, for the lower order moments, stronger than the statistical fluctuations, and are hence a source of systematic error due to finite-$\textrm{Re}$. These fluctuations are an intrinsic property of the dynamical system at hand and have been studied in more detail in \cite{Mailybaev2023}.

\begin{figure*}
    \centering
    \includegraphics[width=0.955\linewidth]{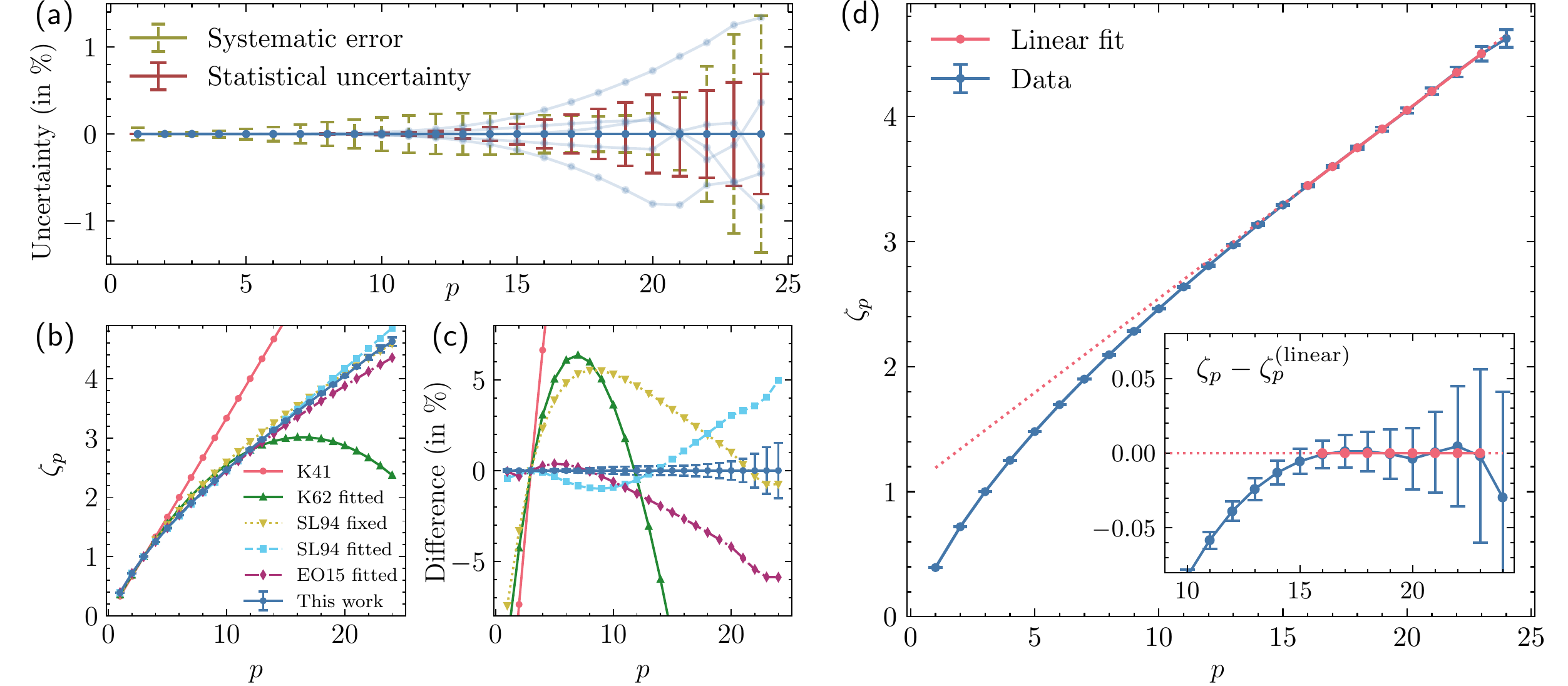}
    \caption{The obtained inertial range scaling exponents $\zeta_p$ as a function of the moment order $p$. Panel a: Relative uncertainty of $\zeta_p$. Dashed olive-green error bars represent the systematic error, while solid red error bars depict the statistical uncertainty. Lines plotted in lighter shade represent realizations of 1/5th subsamples of the total data. Panels b, c: Comparison of the inertial range scaling exponents $\zeta_p$ with various models: Kolmogorov (K41) \cite{Kolmogorov1941}, Lognormal (K62) \cite{Kolmogorov1962}, fitted as $(\mu\simeq0.20)$, She-Leveque (SL94) \cite{She1994} with fixed $(C_0=2, X=2/3)$ and variable $(C_0,X)$, fitted as $(C_0\simeq1.46,X\simeq0.36)$, and Eling-Oz (EO15) \cite{Eling2015}, fitted as $(\gamma^2\simeq0.25)$. While the models capture certain qualitative aspects of the scaling exponents as seen in (b), a quantitative compensated comparison (c) reveals that all models are far outside error bars of the numerical results obtained in this work. Panel d: Comparison of the inertial range scaling exponents $\zeta_p$ with a linear asymptotic scaling. The linear fit is carried out on moments 16 (inclusive) to 24 (exclusive), yielding a slope $\alpha = 0.151 \pm 0.004$ and intersection $\beta = 1.04 \pm 0.06$. It captures the asymptotic scaling of the numerical data and remains within error bars from moment $p\geq 15$ onwards (inset).}
    \label{fig:fig_2}
\end{figure*}

It is thus important that we distinguish and quantify both sources of uncertainty: the uncertainty due to statistical fluctuations from the finite sampling and the systematic error due to finite-$\textrm{Re}$ dissipative oscillations. This uncertainty quantification is treated in Appendix~\ref{app:uncert}. Note that since the systematic error quantifies the spread over the local slopes in the inertial range, this then also accounts for the degree of arbitrariness introduced by the selection of the inertial range.

The resulting estimates of the inertial range scaling exponents $\zeta_p$, and the systematic and statistical uncertainty therein are provided in Fig.~\ref{fig:fig_2}.

\begin{table}[b!]
\caption{\label{tab:zeta_p}
Obtained numerical values of the inertial range scaling exponents $\zeta_p$ and their total uncertainties.}
\begin{minipage}{0.4\linewidth}
\begin{tabular}{cc}
\toprule
& $\zeta_p$ \\
\colrule
$\zeta_{1}$ & $ 0.3932 \pm 0.0003 $\\
$\zeta_{2}$ & $ 0.7197 \pm 0.0002 $\\
$\zeta_{3}$ & $ 1.0001 \pm 0.0002 $\\
$\zeta_{4}$ & $ 1.2504 \pm 0.0005 $\\
$\zeta_{5}$ & $ 1.4805 \pm 0.0009 $\\
$\zeta_{6}$ & $ 1.696 \pm 0.002 $\\
$\zeta_{7}$ & $ 1.900 \pm 0.003 $\\
$\zeta_{8}$ & $ 2.095 \pm 0.003 $\\
$\zeta_{9}$ & $ 2.282 \pm 0.004 $\\
$\zeta_{10}$ & $ 2.463 \pm 0.005 $\\
$\zeta_{11}$ & $ 2.638 \pm 0.006 $\\
$\zeta_{12}$ & $ 2.807 \pm 0.007 $\\
\colrule
\end{tabular}
\end{minipage} 
\begin{minipage}{0.4\linewidth}
\begin{tabular}{cc}
& \vspace{1.6mm} \\
\colrule
$\zeta_{13}$ & $ 2.973 \pm 0.008 $\\
$\zeta_{14}$ & $ 3.134 \pm 0.008 $\\
$\zeta_{15}$ & $ 3.293 \pm 0.009 $\\
$\zeta_{16}$ & $ 3.45 \pm 0.01 $\\
$\zeta_{17}$ & $ 3.60 \pm 0.02 $\\
$\zeta_{18}$ & $ 3.75 \pm 0.02 $\\
$\zeta_{19}$ & $ 3.90 \pm 0.02 $\\
$\zeta_{20}$ & $ 4.05 \pm 0.03 $\\
$\zeta_{21}$ & $ 4.20 \pm 0.03 $\\
$\zeta_{22}$ & $ 4.36 \pm 0.05 $\\
$\zeta_{23}$ & $ 4.50 \pm 0.06 $\\
$\zeta_{24}$ & $ 4.62 \pm 0.08 $\\
\toprule
\end{tabular}
\end{minipage}
\end{table}

As shown in Fig.~\ref{fig:fig_2}(a), indeed, the lower order moments $p\lesssim14$ are dominated by the systematic error from finite-$\textrm{Re}$ oscillations, while the higher order moments $p\gtrsim 14$ are dominated by the statistical error due to finite sampling, as they are governed by increasingly rare events. The resulting values of the scaling exponents $\zeta_p$ and their total uncertainty (combined systematic and statistical uncertainty) are also provided in Tab.~\ref{tab:zeta_p}.

Note that the scaling exponent of the third moment is in agreement with the Kolmogorov 4/5-law that proves $\zeta_3=1$ \cite{Kolmogorov1941,Frisch1995} within the 5th significant digit, giving us confidence that the {\it true} values of other low order exponents are estimated with the same accuracy and put a clear unprecedentedly accurate benchmark for any new theory aiming to rigorously calculate scaling properties in turbulence and in turbulence models.

We confirm furthermore that when increasing the extent of the considered inertial range by one shell on both ends to $5 \leq n < 16$, we obtain $\zeta_3=1.0001 \pm 0.0003$, while decreasing the considered inertial range by one shell on both ends to $7 \leq n < 14$, we find $\zeta_3=1.0001 \pm 0.0002$, underpinning the validity of our results.

The robustness of our results with respect to the resolution, finite $\textrm{Re}$ and the forcing scheme is validated in Appendix~\ref{app:validation}.

\section{Comparison with intermittency models}
With this very precise measurement of the scaling exponents, it is now insightful to compare our numerical results with various intermittency models that have been proposed in the literature.

For completeness, we start by considering the fully self-similar solution predicted by Kolmogorov (K41) \cite{Kolmogorov1941},
\begin{equation}
    \zeta_p^{\textrm{(K41)}}=\frac{p}{3}.
\end{equation}
Historically, also the Lognormal model has been considered, referred to as Kolmogorov `62 \cite{Kolmogorov1962}, which yields a parabolic dependence on $p$ as
\begin{equation}
    \zeta_p^{\textrm{(K62)}}=\frac{p}{3}+\frac{\mu}{18}\left(3p-p^2\right).
\end{equation}
Arguably the most widely celebrated model that incorporates intermittency is the She-Leveque model \cite{She1994}, predicting anomalous scaling exponents given by
\begin{equation}
    \zeta_p^{\textrm{(SL94)}} = \left(1-\frac{C_0}{3}\right) \frac{p}{3} + \frac{C_0}{3} \frac{1}{1-X} \left(1-X^{\frac{p}{3}}\right),
\end{equation}
where the parameter $C_0$ can be interpreted as the co-dimension of the most strongly fluctuating structures, which, in 3D turbulence, are assumed to be 1D vortex filaments, such that $C_0=2$ and $X=2/3$ \cite{She1994}. However, for shell models, these need not be the same, so we consider $C_0$ and $X$ as a free parameters.

An alternative intermittency model was recently proposed by Eling \& Oz \cite{Eling2015}, predicting
\begin{equation}
    \zeta_p^{\textrm{(EO15)}} = \frac{\sqrt{\left(1+\gamma^2\right)^2+4 \gamma^2\left(\frac{p}{3}-1\right)}+\gamma^2-1}{2 \gamma^2},
\end{equation}
with a single free parameter $\gamma^2$.

These models are compared and fitted to the numerical data, weighting with the respective uncertainties to produce a maximum likelihood estimate fit. This comparison is laid out in Fig.~\ref{fig:fig_2}(b,c). While capturing certain aspects of the qualitative shape of $\zeta_p$, the figure shows that the models considered here, at a quantitative level, all fall far outside error bars and are thus not in agreement with our numerical results. For completeness, we also considered the model by Yakhot \cite{Yakhot2001}, the $p$-model by Meneveau \& Sreenivasan \cite{Meneveau1987} and the random $\beta$-model \cite{Benzi1984}, but their agreement (not shown) is less than the models considered above.

Concerning the large $p$ asymptotics, governing the statistics of extreme events, we find that the scaling exponents tend to increase linearly
\begin{equation}\label{eq:asympt}
    \lim_{p\to\infty} \zeta_p = \underbrace{\alpha p + \beta}_{\zeta_p^{\textrm{(linear)}}}.
\end{equation}

Indeed, as shown in Fig.~\ref{fig:fig_2}(d), the numerically obtained exponents follow this linear behavior within error bars from $p\geq 15$ onwards. From linear fitting, we obtain a slope $\alpha = 0.151 \pm 0.004$ and intersection $\beta = 1.04 \pm 0.06$. This suggests that, asymptotically, the high order moments are dominated by one singularity $h\simeq0.15$ in agreement with what was measured with lower precision in \cite{Mailybaev2023,Lvov2001}.We point out, however, that since the linear behavior develops in the region of increasing error bars, irrefutably proving the existence of the asymptotic Eq.~\eqref{eq:asympt} warrants further investigation.


\begin{figure}[b!]
    \centering
    \includegraphics[width=\linewidth]{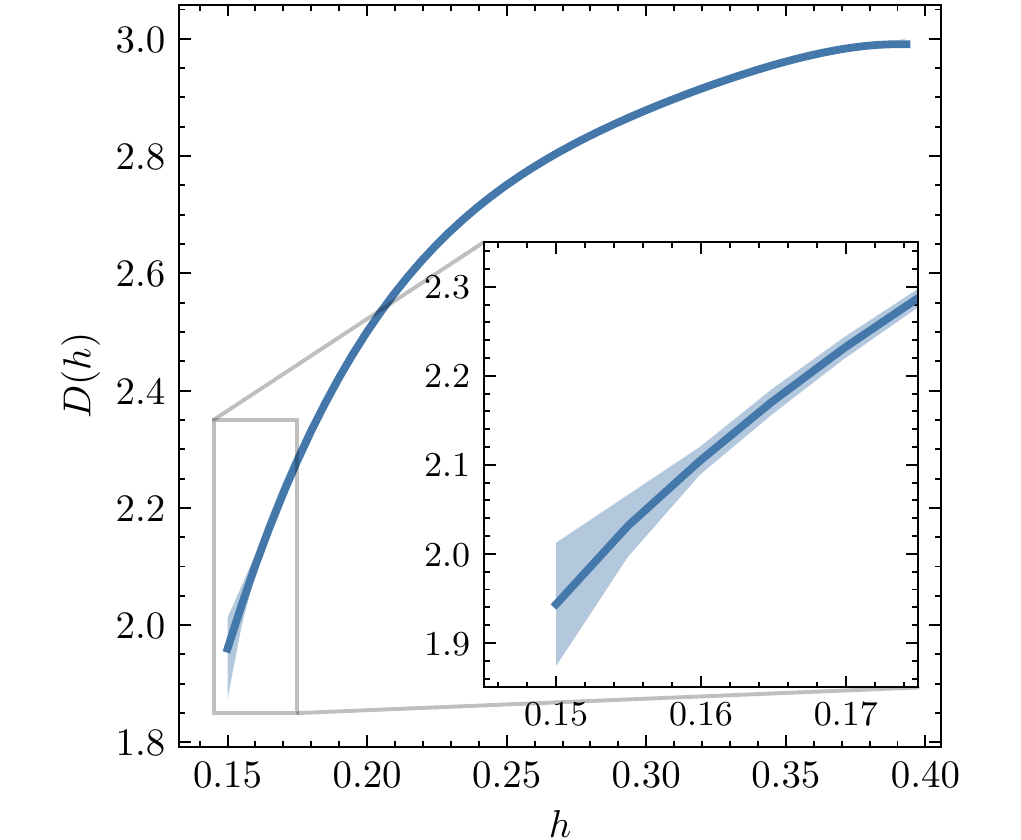}
    \caption{Singularity spectrum $D(h) = 3-C(h)$ of the multifractal description of the shell model intermittency obtained through a Legendre transform of the inertial range scaling exponents $\zeta_p$. The light blue shaded region represents the uncertainty interval, obtained by considering the spread between the minimum and maximum envelope of the uncertainty interval of $\zeta_p$.}
    \label{fig:D_h}
\end{figure}

\section{Singularity spectrum}

Finally, we can interpret our findings in the context of the multifractal model of turbulence \cite{Parisi1985}. In this language, velocity structures scale with variable singularity exponents $h$ as
\begin{equation}
    u_n(t) \sim k_n^{-h} \qquad \forall \quad t\in\mathcal{S}_h,
\end{equation}
with each $h$ between some $h_{\textrm{min}}$ and $h_{\textrm{max}}$ residing in a set $\mathcal{S}_h$ of fractal dimension $D(h)$, with codimension \mbox{$C(h)=3-D(h)$}, which is referred to as the singularity spectrum. The singularity spectrum can then be obtained from the scaling exponents of the structure function $\zeta_p$ through a steepest descent computation, yielding \cite{Parisi1985,Frisch1995}
\begin{equation}
    C(h) = \sup_p \left[\zeta_p - ph\right],
\end{equation}
which shows that the singularity spectrum is connected to the structure function scaling exponents through a Legendre transform.

By computing the Legendre transform on the numerically obtained scaling exponents, we find the singularity spectrum provided in Fig.~\ref{fig:D_h}. The strongest singularity $h_{\textrm{min}}$ is thus indeed provided by the slope of the obtained asymptotic linear scaling $h_{\textrm{min}}\simeq 0.15$, while the intercept can be interpreted as its corresponding co-dimension $C(h_{\textrm{min}})\simeq 1$.

\section{Conclusions and outlook}
By employing modern day GPU acceleration, we have obtained statistics of the intermittency in SABRA shell model for 3D turbulence at unprecedented precision, allowing us to capture moments of the structure function up to $p=24$. Through careful assessment of the different sources of uncertainty in our numerical results, we performed a quantitative comparison with different models for the anomalous scaling exponents describing the turbulence intermittency. This reveals that none of the considered intermittency models \cite{Kolmogorov1941,Kolmogorov1962,She1994,Eling2015,Yakhot2001,Meneveau1987,Benzi1984} are consistent with our numerical results of the shell model turbulence. This need not imply that those models are also falsified in the context of Navier-Stokes turbulence, because there is no proof that Navier-Stokes and shell models scaling must have the same functional shape. Nevertheless, our numerical results define a new benchmark for the precision with which analytical calculations for anomalous scaling in shell models of turbulence must agree with data.

We find that the asymptotic behavior of the inertial range scaling exponents at high order moments $p$ is consistent with a linear scaling. This indicates that the most extreme turbulent events are dominated by one leading singularity that we find to be $h_{\textrm{min}}\simeq 0.15$.

To further improve the fidelity of the current measurement, one would need to consider both sources of uncertainty. To improve the precision at high moments, which are dominated by the statistical uncertainty, even longer time statistics are needed, which will allow one to explore moments even beyond $p=24$. For improvement of the low moments, which are dominated by the systematic error due to structural fluctuations coming from the dissipative range, one needs to consider yet higher $\textrm{Re}$, requiring more shells and a finer time step (scaling with the Kolmogorov time $\tau_\eta\propto\textrm{Re}^{-1/2}$). A promising alternative route to circumvent this systematic error without resorting to prohibitively larger $\textrm{Re}$ is to compute the scaling exponents over the full range of scales in the inertial and dissipative range by using the collapse of the structure functions at two different $\textrm{Re}$ as put forward in \cite{Mailybaev2023}, which this requires a modified dissipative closure of the shell model that is at least of quadratic order. We also mention the existence of a different class of shell models~\cite{mailybaev2021solvable}, which rigorously produce an arbitrary dependence for $\zeta_p$ with linear large-$p$ asymptotics, at the expense of relaxing the energy conservation.

To facilitate and encourage further model development and numerical benchmarking, we make all raw data underlying this work publicly available in Ref. \footnote{Data underlying this work are available on \href{https://zenodo.org/doi/10.5281/zenodo.10611104}{10.5281/zenodo.10611104}.}, which also includes the relevant post-processing routines as well as a notebook that reproduces all figures presented here.

\medskip
\begin{acknowledgments}
We are grateful for the support of the Netherlands Organisation for Scientific Research (NWO) for the use of supercomputer facilities (Snellius) under Grant No. 2307092\_24. We thank Han Verbiesen and Eindhoven University of Technology for granting the computational resources. This publication is part of the project “Shaping turbulence with smart particles” with project number OCENW.GROOT.2019.031 of the research programme Open Competitie ENW XL which is (partly) financed by the Dutch Research Council (NWO). A.A.M. is supported by CNPq grant 308721/2021-7 and FAPERJ grant E-26/201.054/2022. L.B. was supported by the European Research Council (ERC) under the European Union’s Horizon 2020 research and innovation programme Smart-TURB (Grant Agreement No. 882340).
\end{acknowledgments}

\appendix

\begin{figure*}
    \centering
    \includegraphics[width=0.93\textwidth]{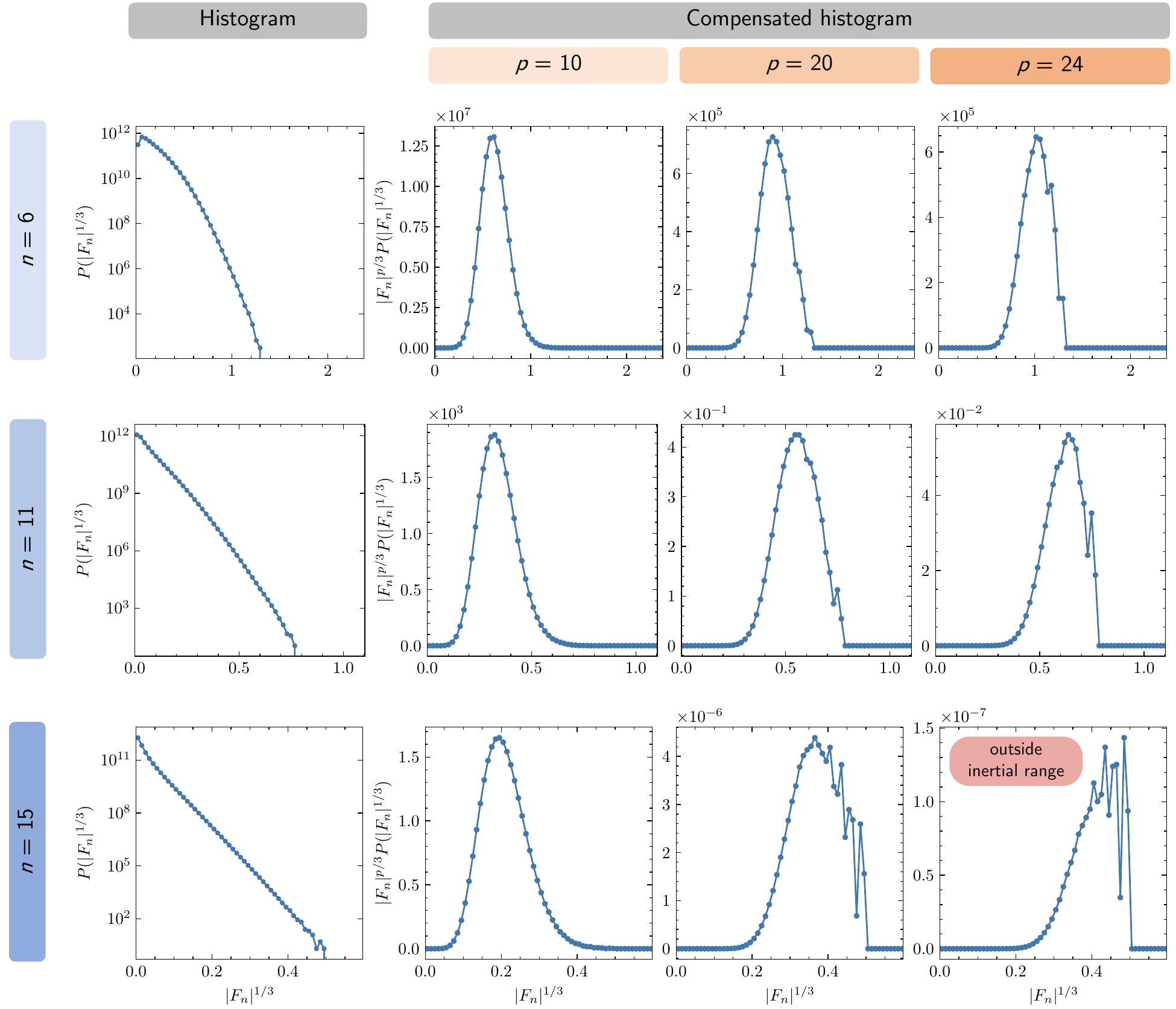}
    \caption{Histograms $P(|F_n|^{1/3})$ (first column) and compensated histograms $|F_n|^{p/3} P(|F_n|^{1/3})$ (second, third and fourth column) of flux structures $|F_n|^{1/3}$. The compensated histograms depict the integrand of the $p$th moment, such that the area under the compensated histogram gives the $p$th moment of the structure function. Columns show the 10th, 20th and 24th moment, respectively. Rows consider different shells at the start of the selected inertial range ($n=6$), inside the inertial range ($n=11$) and at the end of the selected inertial range ($n=15$), respectively.}
    \label{fig:histograms}
\end{figure*}

\section{Compensated histograms}\label{app:hist}

To assess whether we are accurately sampling all considered moments of the structure function, we not only keep track of the accumulated moments, but also tally histograms of each shell in the production code. This gives us access to the probability density $P(|F_n|^{1/3})$ of the flux structures, from which the structure functions follow as
\begin{equation}
    S_p(n) = \langle | F_n |^{p/3} \rangle = \int_0^\infty |F_n|^{p/3} P(|F_n|^{1/3}) \mathrm{d} |F_n|^{1/3}.
\end{equation}
The integrand in the last expression is coined the compensated histogram and from it, we can asses whether the moment is sufficiently well sampled.

(Compensated) histograms of selected shells and moments are provided in Fig.~\ref{fig:histograms}. If the compensated histogram is smooth, this indicates that the moment is well sampled by many significant independent contributions, while a very roughly peaked compensated histogram indicates that the measured moment is dominated by a small number of individual samples and is thus not reliable. We consider the sample of a moment unreliable when spikes of individual bins due to singular events are larger than the smooth maximum of the compensated histogram.

Since higher shells are more intermittent, the higher shells are more difficult to sample (they require increasingly more statistics), which outweighs the fact that their decorrelation time is lower. We have selected the inertial range for the calculation of the anomalous scaling exponents in the main text to be $6\leq n < 15$. 
We find that moderate moments (we show $p=10$) are well resolved over the full inertial range. At moment $p=20$, the last shell in the inertial range $n=15$ is considered just acceptably well sampled. For moments between $20<p\leq24$, we gradually decrement the selected upper bound of the inertial range, because the higher shells are not well enough sampled. We show the highest considered moment $p=24$, for which we consider the inertial range up to shell $n=11$, which is still acceptably well sampled, but higher shells (e.g. $n=15$ as shown) are not reliably sampled as their compensated histogram is dominated by individual contributions.

\section{Uncertainty analysis}\label{app:uncert}

To quantify the statistical uncertainty in our measurement, we compute the scaling exponents $\zeta_p$ over $\mathcal{N}$ equally divided subsamples of the data. With $\zeta_p$ the scaling exponent over the full data and $\hat{\zeta}_{p,i}$ the scaling exponents over the $i$th subsample, we then obtain the statistical uncertainty $\sigma_{\textrm{stat}}$ as the standard error
\begin{equation}
    \sigma_{\textrm{stat}} = \sqrt{\frac{1}{\mathcal{N}(\mathcal{N}-1)}\sum_{i=1}^\mathcal{N} \left( \hat{\zeta}_{p,i} - \zeta_p \right)^2}.
\end{equation}
We have verified that this definition of the statistical uncertainty becomes independent of the choice of number of subsamples $\mathcal{N}$ for any $\mathcal{N}<100$ that we have tested, following the central limit theorem. We arbitrarily choose $\mathcal{N}=5$ in this work. In the main text, we report 95\% (or 2$\sigma_{\textrm{stat}}$) confidence intervals.

To quantify the systematic error owing to the structural fluctuations from the dissipative range (the finite-$\textrm{Re}$ effect, see main text), we consider the spread of the local scaling exponent over the different shells in the inertial range from shell $n_{\textrm{str}}$ to $n_{\textrm{end}}$. This component of the error also accounts for the somewhat arbitrary selection of the inertial range.

We must note, however, that the spread of the local scaling exponents is also convoluted by the statistical fluctuations. To isolate the contributions of systematic errors from the statistical fluctuations, we therefore compensate the spread over the local scaling exponents by the expected statistical fluctuation at the single shell level. This leads to the definition of the systematic error as
\begin{equation}
\begin{aligned}
    \sigma_{\textrm{syst}} = &\sqrt{\frac{1}{\left(n_{\textrm{end}} - n_{\textrm{str}} -1 \right)}\sum_{n_{\textrm{str}} \leq n < n_{\textrm{end}}} \left( \tilde{\zeta}_p(n) - \zeta_p \right)^2} \\
    &- \sigma_{\textrm{stat}} \sqrt{(n_{\textrm{end}} - n_{\textrm{str}})/2},
\end{aligned}
\end{equation}
where we have tacitly assumed that every second shell fluctuates statistically independently.

For the total uncertainty we use a sum of variance
\begin{equation}
    \sigma_{\textrm{tot}} = \sqrt{ \sigma_{\textrm{stat}}^2 +  \sigma_{\textrm{syst}}^2}.
\end{equation}

\section{Validation of resolution, finite-Re and forcing}\label{app:validation}

\begin{figure}[b!]
    \centering
    \includegraphics[width=0.8\linewidth]{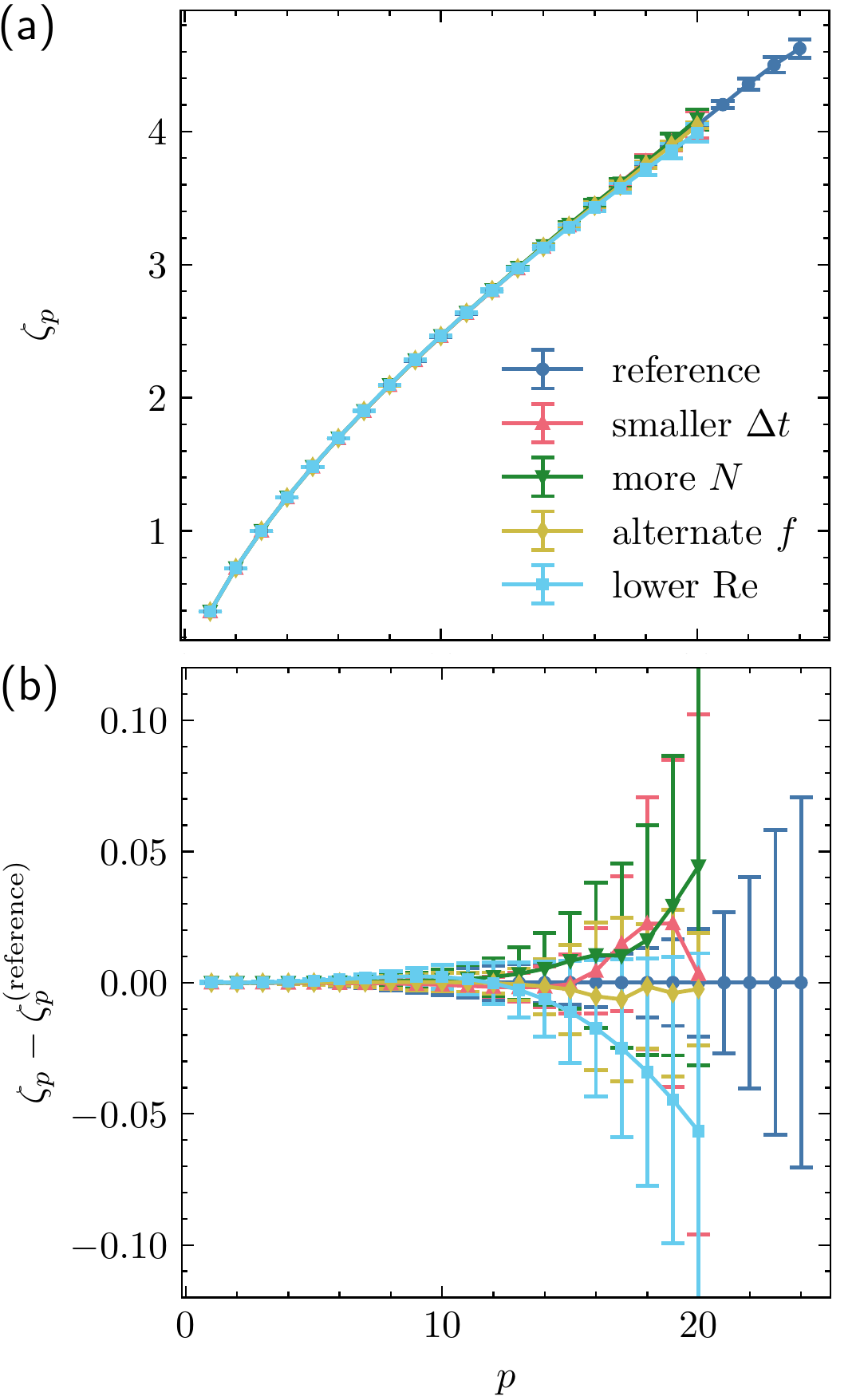}
    \caption{Comparison of the inertial range scaling exponents $\zeta_p$ obtained from validation runs at varying timestep $\Delta t$, number of resolved shells $N$, forcing scheme $f$, and Reynolds number $\textrm{Re}$, to the reference case treated in the main text. The obtained scaling exponents $\zeta_p$ are shown in (a), while they are plotted compensated by the reference case $\zeta_p^{\textrm{(reference)}}$ in (b) to highlight their difference with respect to the error bars.}
    \label{fig:zeta_p_validation}
\end{figure}

To validate the robustness of the obtained results with respect to the temporal and shell-spatial resolution as well as the finite $\textrm{Re}$ and forcing scheme, we perform a quantitative comparison over a limited set of runs where we individually vary the resolutions, $\textrm{Re}$ and change the forcing scheme.

We consider
\begin{equation}
\begin{aligned}
    &\Delta t = 2 \times 10^{-9} = 0.2 \, \Delta t_{\textrm{ref}},\\
    &N=50=N_{\textrm{ref}}+10,\\
    &\textrm{Re}=10^{10}=10^{-2}  \, \textrm{Re}_{\textrm{ref}},
\end{aligned}
\end{equation}
where we use subscript ``$\textrm{ref}$'' to denote the reference values used in the main text. We also consider an alternative deterministic forcing scheme that ensures a fixed energy flux as
\begin{equation}
    f_n = \left( \frac{\epsilon_0}{u_0^* + 2^{-1/3} u_1^* }, \frac{2^{-1/3} \epsilon_0 }{u_0^* + 2^{-1/3} u_1^* }, 0,...,0 \right),
\end{equation}
with $\epsilon_0=0.5$.

The statistics obtained for these validation runs are limited to approximately $9\times 10^5$ large Eddy turnover time (compared to $4\times 10^7$ for the reference production runs in the main text). Therefore, we can only compare up to around moment $p\leq 20$.

The comparison of the obtained inertial range scaling exponents $\zeta_p$ is shown in Fig.~\ref{fig:zeta_p_validation}, showing that all validation runs remain within error bars of the reference case considered in the main text, substantiating the robustness of our results. We note that one should consider the systematic errors at finite $\textrm{Re}$ with some caution, because the law of convergence as $\textrm{Re} \to \infty$ is not known.


%

\end{document}